\documentstyle[epsf,12pt]{article}

\begin{document}
\begin{titlepage}
\centerline{\bf{FINITE TEMPERATURE STUDY OF NUCLEAR MATTER}} 
\centerline{\bf{IN SU(2) CHIRAL SIGMA MODEL}}
\vspace{0.5in}
\centerline { P. K. Jena$^*$ and L. P. Singh$^+$}
\date{}
\vspace{0.25in}
\centerline{Department of Physics, Utkal University, Vanivihar,}
\centerline{ Bhubaneswar-751004, India.}
\vspace{1in}
\centerline{\bf{Abstract}}
\vspace{.1in}

  We study here the equation of state of symmetric nuclear matter at 
finite temperatures using a modified SU(2) Chiral Sigma model.
The effect of temperature on effective mass, pressure, entropy and 
binding energy is discussed. The liquid-gas phase transition is  investigated 
and the critical values of temperature, density and pressure are calculated.

\vspace{0.5in}

 PACS Nos: 26.60+c, 21.30.Fe, 21.60.Jz.

\vspace{1.5in}
\noindent
$^*$email: pkjena@iopb.res.in \\
$^+$email: lambodar@iopb.res.in

\end{titlepage}

\section{Introduction}

  Accurate description of properties of hot, dense nuclear matter is a 
fundamental problem in theoretical physics. To determine the
properties of nuclear matter as functions of density and temperature,
one  must study the ground state and excited
states of normal nuclei along with the highly excited states of normal nuclei 
created in nucleus-nucleus collisions and in nuclei far from stability, which 
may be created in radioactive beams[1]. As the equation of state(EOS) 
describes how the energy 
density and pressure vary with density  and temperature, it is possible to 
study the liquid-gas phase transition, which may occur in the warm and dilute
matter produced in heavy-ion collisions. Several authors had studied the 
liquid-gas phase transition using non-relativistic theory[2] and the critical
temperature has been estimated to be in the range 15-20 MeV. But the
experimental results for warm 
and dilute nuclear matter produced in heavy-ion collisions support a small 
liquid-gas phase region with a low critical temperature $T_c \approx $ 13.1
$\pm $ 0.6 MeV[3]. For symmetric nuclear matter, using relativistic mean field 
theory, the critical temperature given by Walecka model is  $T_c \approx$ 18.3
MeV which reduces to $T_c \approx$ 14.2 MeV[4] if the non linear terms are 
included.

  In the present paper, we continue our investigation with modified SU(2) 
chiral sigma model(MCH)[5,6] using relativistic mean-field calculations for 
symmetric hot, dense nuclear matter. The earlier studies have SU(2) chiral 
sigma model as a viable microscopic model for description of dense nuclear 
matter. In the present work, We study the effect of temperature on the EOS,
effective mass, entropy and binding energy for symmetric nuclear matter and
also investigate the liquid-gas phase transition.

   This paper is organized as follows: In sec.2, we present nuclear matter at 
finite temperature. The results and discussion are presented in sec.3. We end
with concluding remarks in sec.4.

\section{\bf{Nuclear matter at finite temperature}}

    We use MCH[5,6] model for symmetric nuclear matter since chiral 
model has been very successful, as such, in describing high density 
nuclear matter. The importance of chiral symmetry[7] in the study of 
nuclear matter was first emphasized by Lee and Wick[8]  and has become 
over the years, one of most useful tools to study high density nuclear matter 
at the microscopic level.     
The nonlinear terms in the chiral sigma model give rise to the three-body 
forces which become significant in the high density regime[9]. Further, 
the energy per nucleon at saturation needed the introduction of
isoscalar field[10] in addition to the scalar field of
pions[11]. The MCH model[5,6] considered by us   
includes two extra higher order scalar field interaction terms 
which ensures an appropriate incompressibility of symmetric nuclear 
matter at saturation density. Further, the equation of state(EOS)
derived from this model is compatible with that inferred from recent 
heavy-ion collision data[12].

   The EOS for hadronic phase is calculated by using the  Lagrangian 
density[5] (with $\hbar=c=K_{Boltzmann}=1$),
\begin{eqnarray}
 L = \frac{1}{2}(\partial_{\mu} \vec{\pi}.\partial^{\mu}\vec{\pi} + 
   \partial_{\mu}\sigma  
  \partial^{\mu}\sigma )-\frac{1}{4}F_{\mu \nu} F_{\mu \nu}-
   \frac{\lambda}{4}(x^2-x_0^2)^2 
  -\frac{\lambda B}{6m^2}(x^2-x_0^2)^3 \nonumber \\ 
   -\frac{\lambda C}{8m^4}(x^2-x_0^2)^4 - g_{\sigma}\bar{\psi }(\sigma +
  i\gamma_{5}\vec{\tau} .\vec{\pi} )\psi  
 +\bar{\psi}(i \gamma_{\mu}
  \partial ^{\mu} -g_{\omega}\gamma_{\mu}\omega ^{\mu})\psi   
   +\frac{1}{2}g_{\omega}^2  x^2 \omega_{\mu}\omega ^{\mu} 
\end{eqnarray}
\noindent
In the above Lagrangian, $F_{\mu \nu} \equiv \partial_{\mu}\omega_{\nu} 
 - \partial_{\nu} \omega_{\mu}$ and $x = (\vec{\pi}^2 +\sigma ^2)^{1/2}$, 
$\psi $ is the nucleon  isospin doublet, $\vec{\pi}$ is the 
pseudoscalar-isovector pion field, $\sigma$ is the scalar field and 
$\omega_{\mu}$, is a dynamically generated isoscalar vector field, which
couples to the conserved baryonic current
$j_{\mu}=\bar{\psi}\gamma_{\mu}\psi$. B and C are constant  coefficients 
associated  with the higher order self-interactions of the scalar field.

   The masses of the nucleon, the  scalar meson  and the vector meson
are respectively given by 
\begin{equation}
  m = g_{\sigma}x_0, \  m_{\sigma} =\sqrt{2\lambda} x_0, \ 
 m_{\omega} = g_{\omega}x_0 
\end{equation}
\noindent
 Here $x_0$ is the vacuum expectation value of the  $\sigma $ field , 
$ g_{\omega}$ and $g_{\sigma}$
are the coupling constants for the vector and scalar fields respectively  
 and $\lambda =
(m_{\sigma}^2 - m_{\pi}^2)/ (2 \it {f}_{\pi}^2) $, where  $m_{\pi}$ is the 
pion mass , $\it{f}_{\pi}$  is the pion decay coupling constant .

   Using Mean-field  approximation, the equation of motion for vector meson 
 field  is 
\begin{equation}
   \omega_{0}=\frac{n_{B}}{g_{\omega}x^2}
\end{equation}
\noindent 
$n_B$ is the baryon number density at temperature T 
and is given by
\begin{equation}
   n_B = \frac{\gamma}{(2\pi)^3}\int_{0}^{\infty} d^3k [ n(T) - \bar n(T)]
\end{equation}
\noindent 
with $n(T) = \frac{1}{e^{( E^*(k) - \nu )\beta} + 1}$ ,\ \ \ \ 
  $ \bar n(T) = \frac{1}{e^{( E^*(k) + \nu )\beta} + 1}$ ,\\
where $E^*(k) = (k^2+y^2m^2)^{1/2}$, the effective nucleon energy, 
$\nu = \mu_{B} - C_{\omega}n_B/y^2$, the effective baryon chemical potential, 
$\beta = 1/K_BT$, 
$\gamma$ is the spin degeneracy factor  with n(T) and $\bar n(T)$ being 
Fermi-Dirac distribution functions for particle and anti-particle at finite
temperature and $y \equiv x/x_0 $ is the effective mass factor which must be 
determined self consistently from the equation of motion for scalar field 
which is given by
\begin{equation}
 (1-y^2)-\frac {B}{m^2C_{\omega}}(1-y^2)^2 +\frac {C}{m^4C_{\omega}^2}
     (1-y^2)^3 +\frac{2 C_{\sigma}C_{\omega}n_{B}^2}{m^2y^4}
   -\frac{C_{\sigma}\gamma}{\pi^2} 
     \int_{0}^{\infty} \frac{dk k^2 (n + \bar n)}{\sqrt{k^2+{m^*}^2}} = 0
\end{equation}
\noindent
 $m^* \equiv ym$ is the effective mass of the nucleon and the coupling 
constants are 
\begin{equation}
   C_{\sigma} \equiv  \frac {g_{\sigma}^2}{m_{\sigma}^2},\ \ 
   C_{\omega}\equiv \frac {g_{\omega}^2}{m_{\omega}^2} \nonumber
\end{equation} \nonumber
\noindent
The energy density and pressure at finite temperature and finite density are 
given by
\begin{eqnarray}
   \epsilon= \frac{m^2(1-y^2)^2}{8C_{\sigma}}-\frac{B}{12C_{\omega} C_{\sigma}}
 (1-y^2)^3 +\frac{C}{16m^2C_{\omega}^2C_{\sigma}}(1-y^2)^4 \nonumber\\
+\frac{C_{\omega}n_B^2}{2y^2}+
   \frac{\gamma}{2\pi^2} \int_{0}^{\infty} dk 
   k^2\sqrt{(k^2+{m^*}^2)} (n + \bar n)
\nonumber \\ 
  P= -\frac{m^2(1-y^2)^2}{8C_{\sigma}} +\frac {B}{12C_{\omega} C_{\sigma}}
   (1-y^2)^3-\frac{C}{16m^2C_{\omega}^2C_{\sigma}}(1-y^2)^4 \nonumber\\
 +\frac{C_{\omega}n_B^2}{2y^2}+
  \frac{\gamma}{6\pi^2}\int_{0}^{\infty}
   \frac{dk k^4(n + \bar n)}{\sqrt{(k^2+{m^*}^2)}}
\end{eqnarray}
\noindent
The entropy density can be obtained as 
\begin {equation}
 S = \frac{1}{T}\left[\frac{C_\omega n_B^2}{y^2} + \frac{\gamma}{2\pi^2}
    \int_{0}^{\infty} dk k^2 \sqrt{(k^2+{m^*}^2)} (n + \bar n)  
+ \frac{\gamma}{6\pi^2}\int_{0}^{\infty}
   \frac{dk k^4(n + \bar n)}{\sqrt{(k^2+{m^*}^2})}\right] -\frac{\mu_B n_B}{T}
\end{equation} 
\noindent

   The values of four parameters $C_\sigma, C_{\omega} $, B and
C occurring in the above equations are obtained by fitting with the 
saturation values of binding energy/nucleon 
(-16.3 MeV), the saturation density (0.153 fm$^{-3}$),
the effective(Landau) mass (0.85M)[13], and nuclear incompressibility 
($\sim $300 MeV), in accordance with recent heavy-ion  collision
data[12] are $C_{\omega}$ = 1.999 fm$^2$, $C_{\sigma}$ = 6.816
fm$^2$, B = -99.985 and C = -132.246.

   Thus the energy density, pressure and entropy density for symmetric nuclear 
matter at finite temperature and finite density can be evaluated by first
solving equations (4) and (5) self consistently for fixed choosen
value of the inverse temperature $\beta$ with variation of  $n_B$ to
get $\mu_B$ and y and then making substitution in equations (7) and (8).

\section{\bf{Results and discussions }}

   In Fig.1, we show pressure as a function of density $n_B$ at different 
temperatures. For a given $n_B$, the pressure has the usual trend of 
increasing  with temperature[4]. As the temperature increases the EOS 
becomes stiffer. Pressure has a non-zero value for $n_B = 0$ at and
above  a temperature of 200 MeV. It indicates that pressure has a 
contribution arising from the thermal distribution functions for
baryons  and anti-baryons as well as from the non-zero
value of the scalar field. Similar results were also obtained by Panda 
$\it{et \ al.}$[14] for symmetric nuclear matter. The non-zero value 
for scalar $\sigma$ field has also been observed in Walecka model[4].
\begin{figure}[t]
\leavevmode
\protect\centerline{\epsfxsize=5in\epsfysize=5in\epsfbox{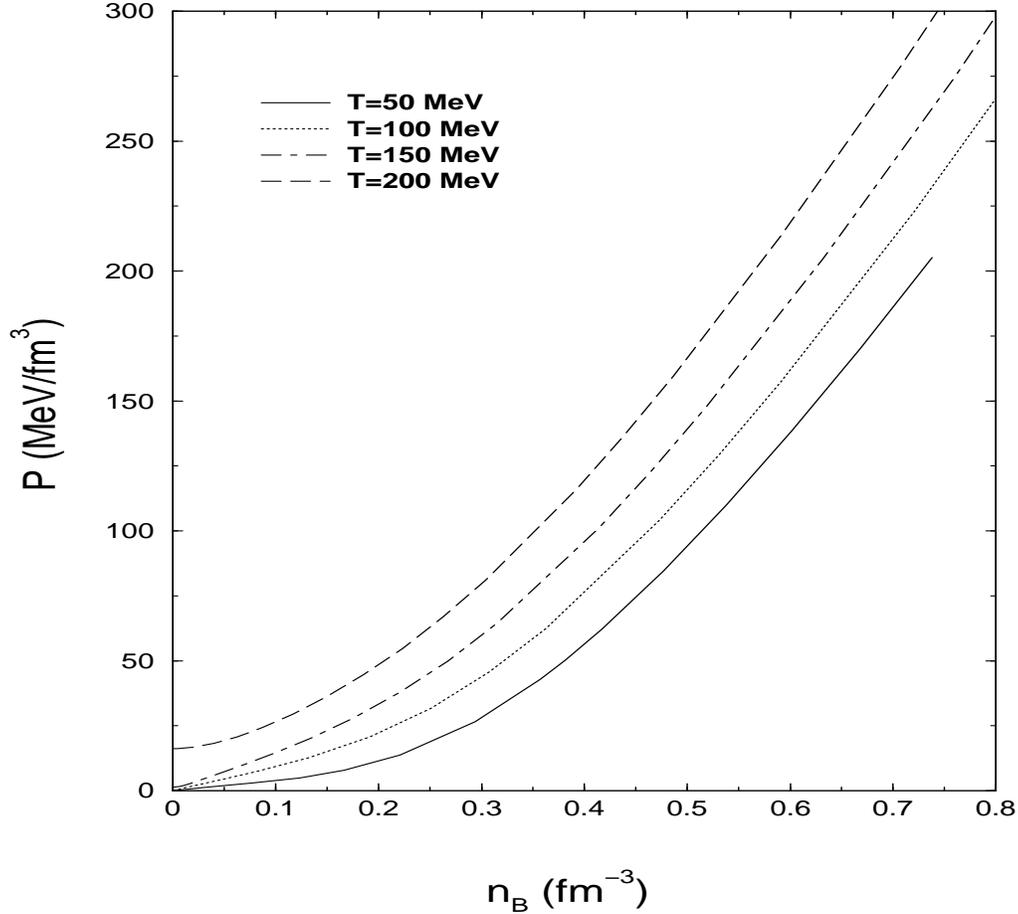}}
\caption{\it{Pressure (P) as function of  baryon  number density
 at different temperatures.}} 
\end{figure}

    Fig.2 shows the behavior of the effective nucleon mass with density at 
different temperatures. We observe that for low temperatures the results are 
not so different from those obtained at zero temperature. This implies that
for low temperatures, the density dependence is more important than the 
temperature dependence. At 200 MeV the effective nucleon mass $m^*$ first 
increases and then decreases slowly. This behavior indicates that the 
effective nucleon mass is considerable in contrast to the Walecka model[15] but
quite similar to that in ZM models[16].
\begin{figure}[t]
\leavevmode
\protect\centerline{\epsfxsize=5in\epsfysize=5in\epsfbox{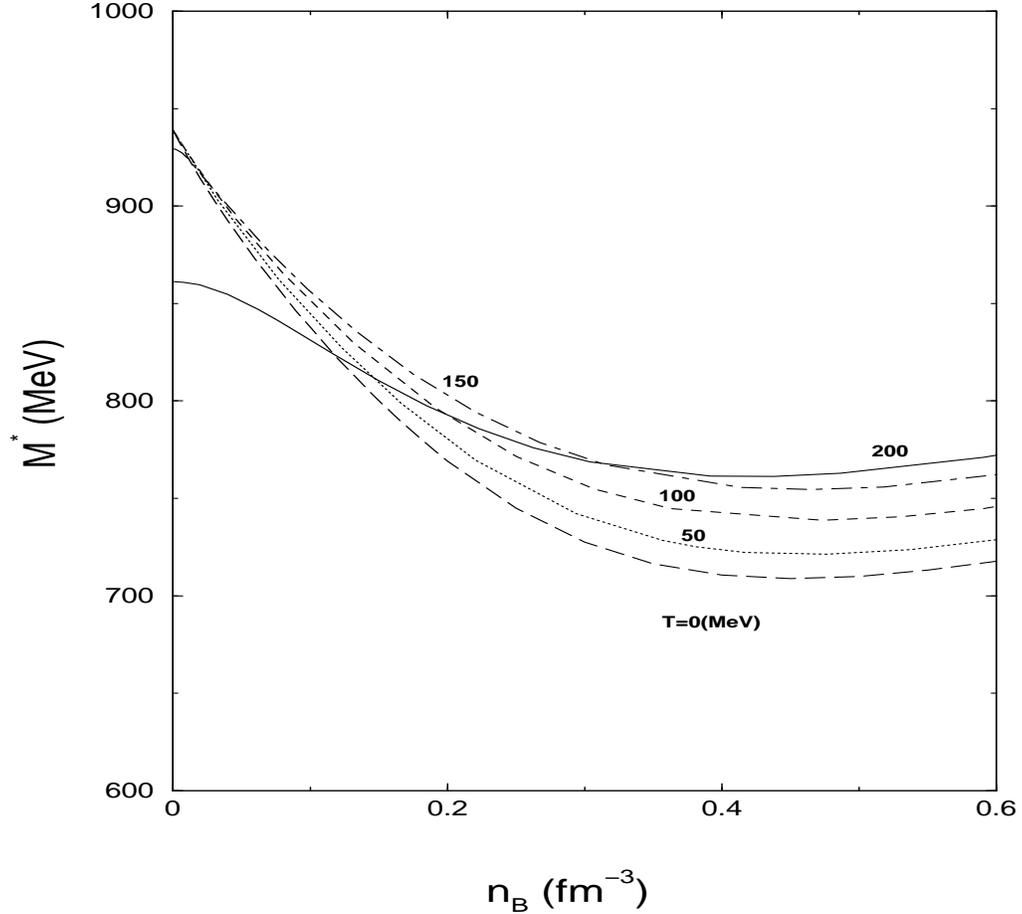}}
\caption{\it{Effective mass as function of  baryon  number density
 at different temperatures.}} 
\end{figure}

    In Fig.3 we show the effective mass as a function temperature at  zero 
density($n_B$ = 0) and saturation density($n_B$ = 0.153fm$^{-3}$).
It is observed that for $n_B = n_0$, the $M^*$ first increases slowly
and then falls suddenly at  
about T $\approx$ 240 MeV. But for $n_B = 0$, $M^*$ remains almost constant as 
the temperature increases and falls suddenly at about T $\approx$ 235 MeV. This
result shows that a first order phase transition appears at $n_B = 0$, 
T $\approx$ 235 MeV which is similar to the result obtained for Walecka model, 
which has such a phase transition at T $\approx$ 185 MeV[17]. Because of 
strong strong attraction between the nucleons at  high temperatures the 
nucleon antinucleon pairs can be formed which may lead to abrupt change 
in $M^*$ to take place in high temperature region.  
\begin{figure}[t]
\leavevmode
\protect\centerline{\epsfxsize=5in\epsfysize=5in\epsfbox{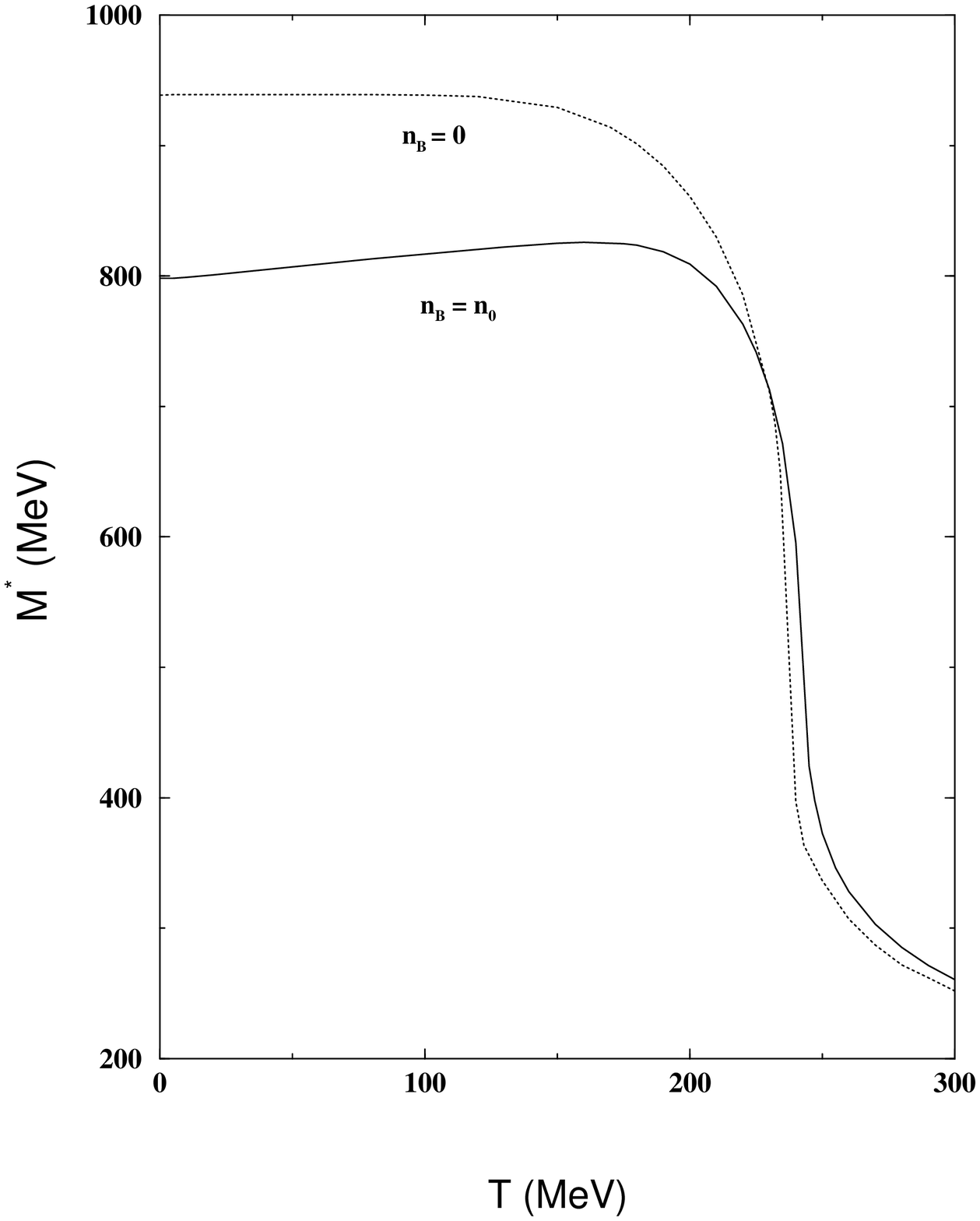}}
\caption{\it{Effective mass as function of temperature for constant
  baryon number densities.}} 
\end{figure}

     The entropy density as a function of density at different temperatures is
presented in Fig.4. It is observed that entropy density is non-zero even at
vanishing baryon density at and above a temperature of 150 MeV with 
contributions from the non-zero value of the sigma field. Similar behavior
was also observed for entropy density in the Walecka model and ZM model
calculations[16]. This increase of entropy density with increase of temperature
indicates a phase transition.
\begin{figure}[t]
\leavevmode
\protect\centerline{\epsfxsize=5in\epsfysize=5in\epsfbox{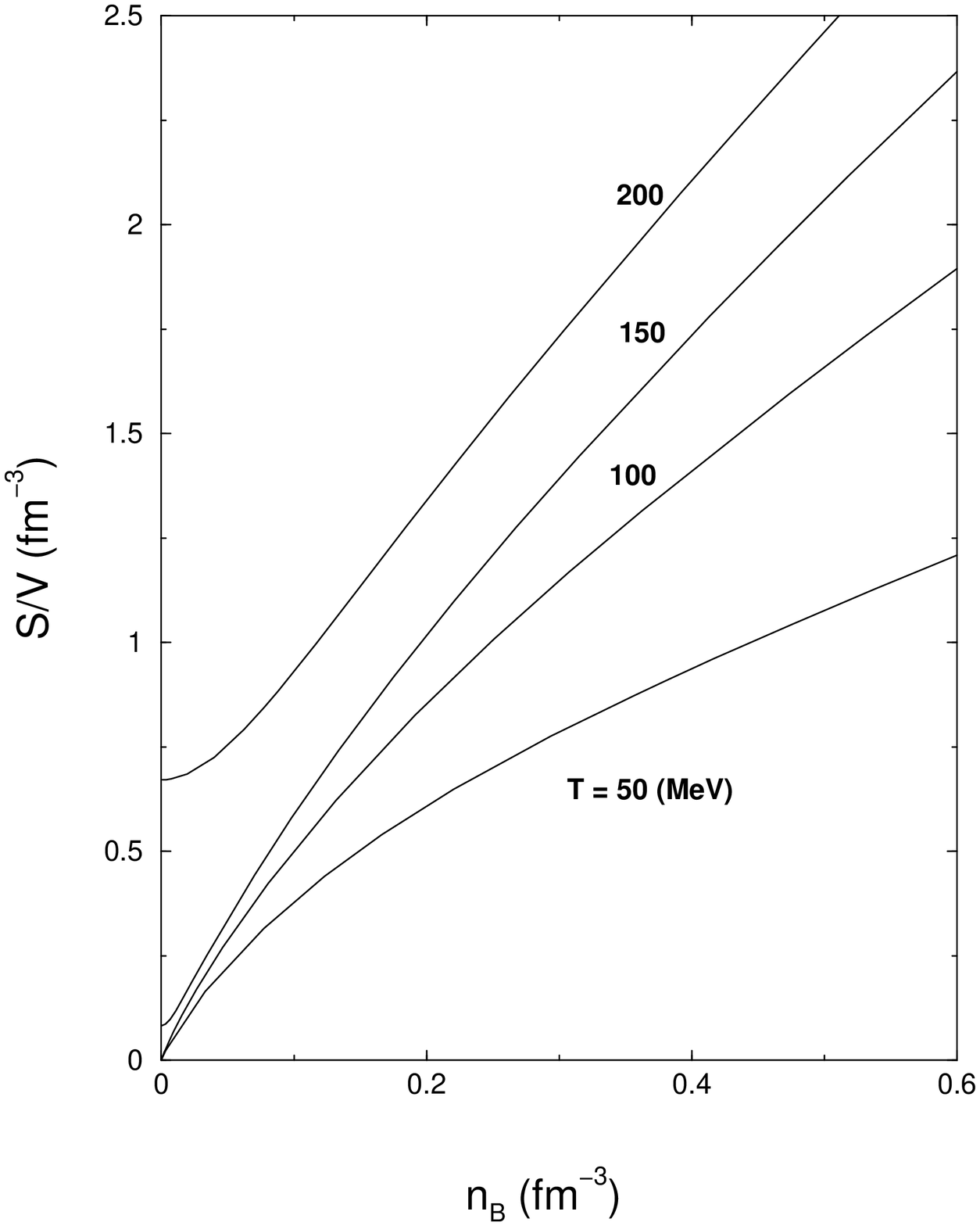}}
\caption{\it{Entropy density as function of  baryon  number density
 at different temperatures.}}
 \end{figure}

     Fig.5 shows the energy per nucleon as a function of the baryon density at
different temperatures. With the increase of temperature the minimum
shifts towards higher densities and for higher temperatures the
minimum of the curve becomes positive. It is also observed that as 
the temperature increases, the nuclear matter 
becomes less bound and the saturation curves in the MCH model are flatter 
than those observed in Walecka model[15,16]. This result implies that the
nuclear matter EOS in MCH model is softer than that obtained in Walecka model.
\begin{figure}[t]
\leavevmode
\protect\centerline{\epsfxsize=5in\epsfysize=5in\epsfbox{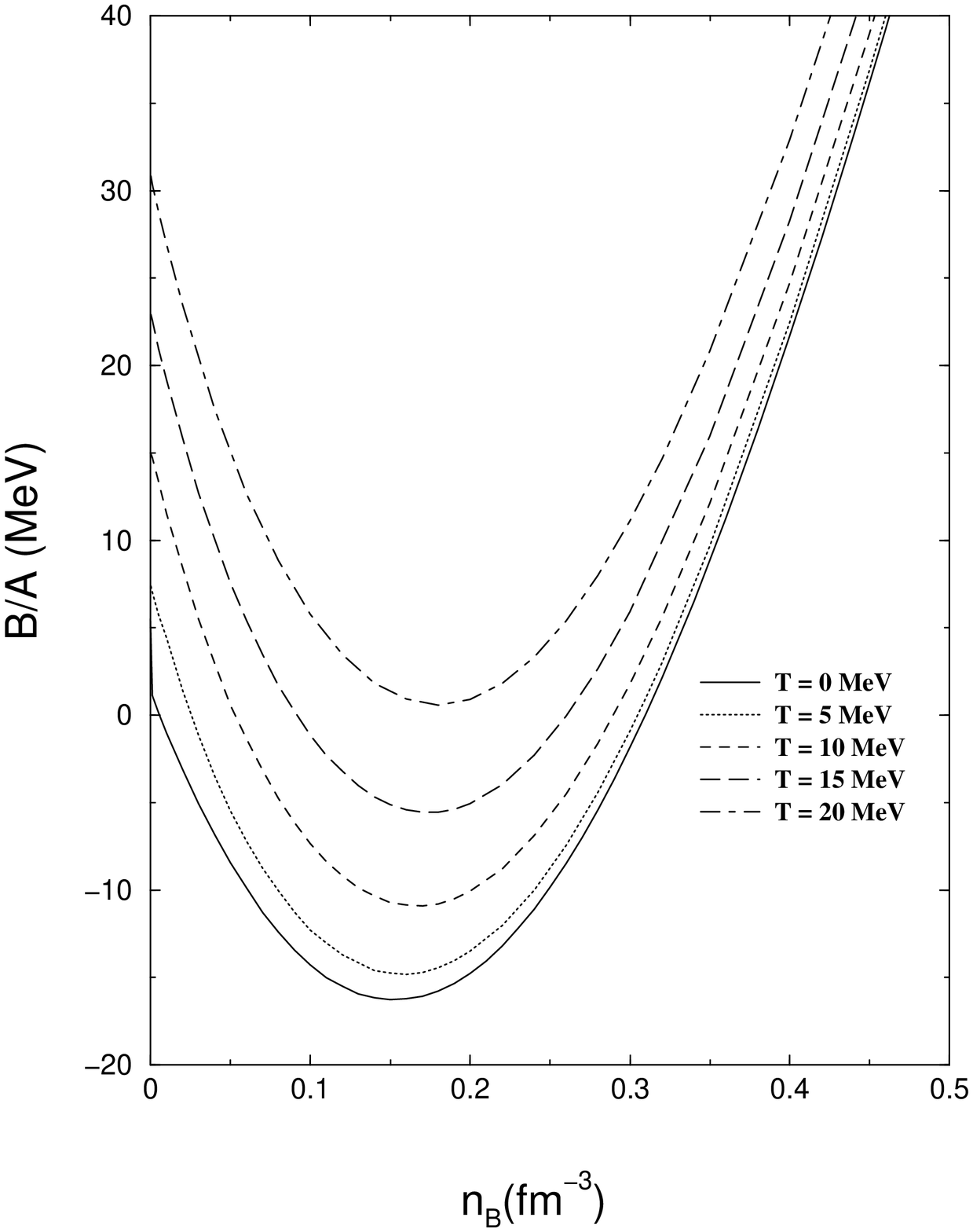}}
\caption{\it{Energy per nucleon as function of  baryon  number density
 at different temperatures.}}
\end{figure}

     Lastly  we present the pressure of nuclear matter as a function of baryon
density for low temperatures in Fig.6. The figure shows that at zero
temperature, the pressure first decreases, then increases and passes through 
P = 0 at $n_B = n_0$(saturation density), where the binding energy per nucleon
is a minimum. Decrease of pressure with density implies a negative 
incompressibility, $K = 9(\frac{\partial P}{\partial n_B})$, which is a
 sign of mechanical instability. When the temperature increases the region
 of mechanical instability decreases and disappears at the critical 
temperature $T_c$, which is determined by $\frac{\partial P}
{\partial n_B}\mid_{T_c} = \frac{\partial^2 P}{\partial^2 n_B}\mid_{T_c} = 0$,
above which the liquid-gas phase transition is continuous. 
We have obtained the value of critical temperature $T_c \approx 17.2
MeV$, critical density  $n_c \approx 0.045 fm^{-3} $,  
critical pressure $ p_c \approx 0.274 MeV/fm^{-3}$ and which is in fair 
agreement with the results obtained in other studies[15,16,18]
\begin{figure}[t]
\leavevmode
\protect\centerline{\epsfxsize=5in\epsfysize=5in\epsfbox{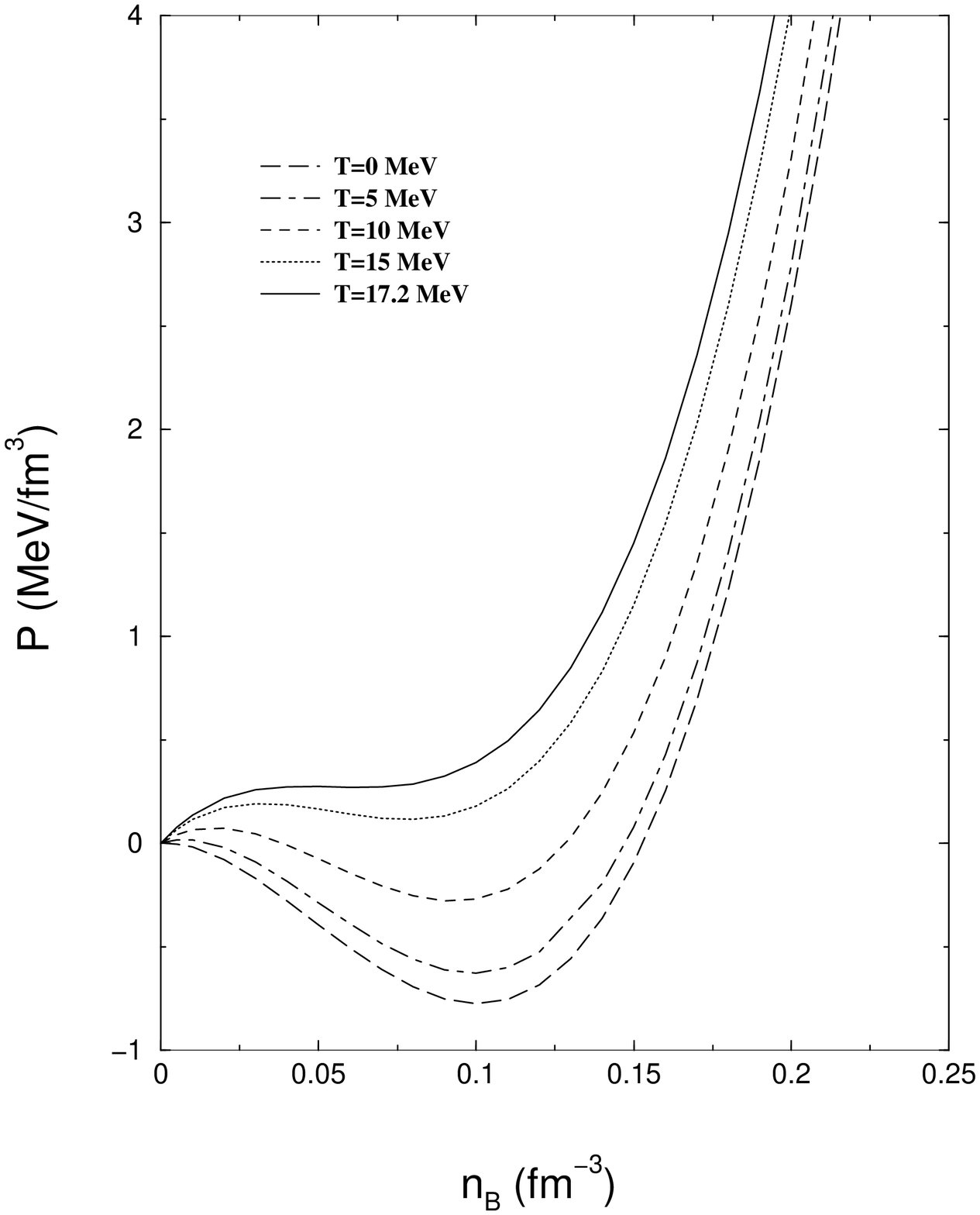}}
\caption{\it{Pressure (P) as function of  baryon  number density
 at different temperatures.}}
\end{figure}

\section{\bf{Conclusions}}

    In this work, we have extended our earlier[6] investigations with     
MCH model at zero temperature to study 
the thermodynamic properties of symmetric nuclear matter at finite temperature.
We have presented variations of pressure, effective nucleon mass, entropy 
density and energy per nucleon with respect to density for various 
temperatures. At zero  density it is found that  
this model exhibit a phase transition at T$\approx$ 235 MeV just as 
obtained in the Walecka model at T$\approx$ 185 MeV[17].
We find in the zero temperature case, the MCH model gives a 
softer EOS of nuclear matter at finite temperature than the Walecka model.
We also find that the model under investigation indicates a liquid-gas phase
transition and the critical temperature is found to be $T_c\approx $ 17.2 MeV
above which the liquid gas phase transition is continuous. The fair
amount of agreement of our results with those of similar studies undertaken
with different models strongly support MCH model as a viable model for
description of nuclear matter at high density.

\vspace {0.1in}
\noindent {\bf{Acknowledgements}}
\vspace {0.05in}

   P.K.Jena would like to thank Council of Scientific and Industrial
Research, Government of India, for the award of 
SRF, F.No. 9/173 (101)/2000/EMR-I. We are  also thankful to Institute
of Physics, Bhubaneswar, India, for providing the library and
computational facility. 



\begin{thebibliography}{99}
\bibitem{ref1}H. M\"{u}ller and B. D. Serot, Phy. Rev. C $\bf{52}$, 2072(1995).
\bibitem{ref2}W. A. Kupper, G. Wegmann and E. R. Hilf,
  Ann. phys.(N.Y.) $\bf{88}$, 454(1974); 
 B. Freedman  and V. R. Pandharipande, Nucl. phys. A  $\bf{361}$, 502(1981); 
 H. Jaqaman, A. Z. Mekjian and L. Zamick, Phy. Rev. C $\bf{27}$, 2782(1983);
 D. Bandyopadyay, C. Samanta, S. K. Samaddar and J. N. De, 
   Nucl. phys. A  $\bf{511}$, 1(1990).
\bibitem{ref3}T. Li $\it{et. al}$, Phy. Rev. C  $\bf{49}$, 1630(1994).
\bibitem{ref4}R. J. Furnstahl and B. D. Serot, Phys. Rev.C $\bf{41}$, 
  262(1990).
\bibitem{5} P. K. Sahu and A. Ohnishi, Prog. Theor. phys.
  $\bf{104}$, 1163 (2000).
\bibitem{ref6} P. K. Jena and L. P. Singh, Mod. Phys. Lett. A   
   $\bf{17}$, 2633 (2002); P. K. Jena and L. P. Singh, Mod. Phys. Lett. A   
   $\bf{18}$, 2135 (2003). 
\bibitem{ref7}M. Gell-Mann and M. Levy, Nuovo Cim. $\bf{16}$, 705 (1960).
\bibitem{ref8}T. D. Lee and G. C. Wick, Phys. Rev. D  $\bf {9}$, 2241 (1974).
\bibitem{ref9}A. D. Jackson, M. Rho, Nucl. Phys. A  $\bf{407}$, 495 (1985).
\bibitem{ref10}J. Boguta, Phys. Lett. B $\bf{120}$, 34 (1983);
  $\bf{128}$, 19 (1983).  
\bibitem{ref11} P. K. Sahu, R. Basu and B. Datta, Astrophys.J. $\bf
  {416}$, 267 (1993).
\bibitem{ref12}P. K. Sahu, A. Hombach, W. Cassing, M. Effenberger and
  U. Mosel, Nucl.Phys. A $\bf{640}$, 493 (1998) ;
   P. K. Sahu, W. Cassing, U. Mosel and A. Ohnishi,  
   Nucl. Phys. A  $\bf {672}$, 376 (2000) ;
    P. K. Sahu, Phys. Rev. C  $\bf{62}$, 045801 (2000). 
\bibitem{ref13}P.M$\ddot{o}$ller, W.D.Myers, W.J.Swiatecki and
  J.Treiner,  Atomic \ Data \ Tables  $\bf{39}$, 225 (1988).
\bibitem{ref14}P. K. Panda, A. Mishra, J. M. Eisenberg and W. Greiner
  Phys. Rev. C  $\bf{56}$, 3134(1997).
\bibitem{ref15}Guo Hua, Liu Bo, M. Di Toro, Phys. Rev. C $\bf{62}$, 
  035203(2000).
\bibitem{ref16} M. Malheiro, A. Delfino and C. T. Coelho, Phys. Rev. C 
  $\bf{58}$, 426(1998). 
\bibitem{ref17} J. Theis, H. St$\ddot{o}$cker  and J. Polonyi, Phys.Rev.D   
 $\bf{28}$, 2286 (1983).
\bibitem{ref18}P. K. Panda, G. Krein, D. P. Menezes and C. Providencia, 
   Phys. Rev. C $\bf{68}$, 015201(2003). 
\end{thebibliography}
\end{document}